# Music Generation Using an LSTM

Michael Conner, Lucas Gral, Kevin Adams

David Hunger, Reagan Strelow, and Alexander Neuwirth

Department of Electrical Engineering and Computer Science

Milwaukee School of Engineering

{connerm, grall, adamsk, hungerd, strelowr, neuwirtha}@msoe.edu

## Abstract

Over the past several years, deep learning for sequence modeling has grown in popularity. To achieve this goal, LSTM network structures have proven to be very useful for making predictions for the next output in a series. For instance, a smartphone predicting the next word of a text message could use an LSTM. We sought to demonstrate an approach of music generation using Recurrent Neural Networks (RNN). More specifically, a Long Short-Term Memory (LSTM) neural network. Generating music is a notoriously complicated task, whether handmade or generated, as there are a myriad of components involved. Taking this into account, we provide a brief synopsis of the intuition, theory, and application of LSTMs in music generation, develop and present the network we found to best achieve this goal, identify and address issues and challenges faced, and include potential future improvements for our network.

# 1 Introduction

Breakthroughs in artificial intelligence have created a variety of possibilities in the recognition, creation, and analysis of music through computational systems. To advance the expanding field, we sought to construct a network built from LSTMs which generates a sequence of pitches and durations which will construct a full song.

The first attempt of computer-generated music was recorded in the 1950s when Alan Turing and Manchester Mark II focused on algorithmic music creation, opening the door to a multitude of possibilities in music intelligence research. Using mathematical models and the Monte Carlo algorithm, which converts randomly generated numbers into musical features such as pitch and rhythm, Illiac Suite for String Quartet, by Lejaren Hiller and Leonard Isaacson, was the first successful production of artificial generation. Their achievements proved that advanced artificial systems are capable of recognizing, creating, and analyzing music. [7]

In the 2017 article post [6] "How to Generate Music using a LSTM Neural Network in Keras" by Sigurður Skúli, the machine learning engineer created music through the usage of an deep neural network in Python using the Keras library. Enlisting a mapping function and designing a model architecture implementing four-layer types, LSTM, dropout, dense, and the activation layer, we used the structure of the network presented here as a springboard for our work.

Although there are a multitude of methods that apply to artificially generating music, architecture design remains nontrivial, training parameters remain poorly understood, and issues of overfitting and ambiguity in the most effective method to model complex data features abound. Addressing these issues, we adapted the music data format into a compatible shape, beginning with a base open-source model which generates quarter notes within a single octave, and trained our neural network on primarily single-track audio with a consistent musical style.

This paper presents the application of an LSTM network for music generation, includes a brief, comprehensible overview into the intuition and theory behind LSTMs and their application in music sequence modeling, provides an analysis on the benefits and disadvantages of our network and training data, shows the qualitative impact on our model output, and discusses potential improvements for our network.

# 2 Methods and Project Pipeline

## 2.1 RNNs

Imagine that you were to start humming along to a song on the radio. This task would be impossible if you were only allowed to hear a single note of the song. However, given some time to listen, you would much more easily be able to pick up and sing along. Likewise, a traditional neural network is unable to remember information that is given over a period of time. This is not just the case for music, but for any sequential information where the next piece of information depends on what happened before. To solve this limitation of traditional neural networks, RNNs (recurrent neural networks) can be applied to information that comes in a sequence. Simple RNNs only have one difference from traditional neutral networks: they have an output that feeds back into the network. [3]

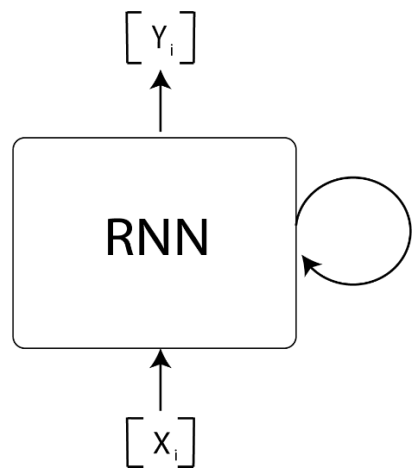

*An RNN cell*

The looping output is a vector which encodes the state of all the network's previous inputs. As the outputs continue to loop, the network essentially has a record of all the previously received information. [3] The looping nature of RNNs allows state about a sequence to be carried from one input to the next. Unraveling this loop gives a clearer picture of how RNNs remember information to generate a sequence. [3]

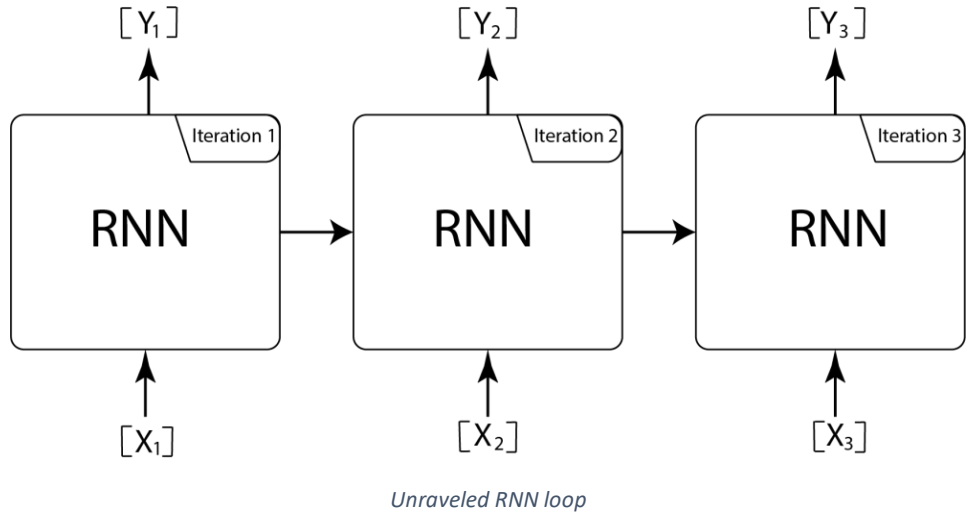
*Unraveled RNN loop*

Each rectangle represents different iterations of the same RNN. The horizontal channel of arrows represents the looping state of the RNN being passed from one iteration to the next.

An RNN could be thought of as a regular neural network with the addition of the network's input being the item of a sequence and with outputs dedicated to encoding and remembering state about the previous items sent through the network.

## 2.2 LSTMs – Theory and Background

A simple RNN might only consist of a single layer dedicated to both remembering a sequence and calculating the output. The reason we didn't use a simple RNN for our project is because of an undesirable property of simple RNNs where previous inputs 'disappear' as more inputs are fed to the network. [3] This is known as the 'vanishing gradient problem,' and it causes previous (and usually important) inputs to influence the output less if they were seen too far in the past. [4] LSTMs were made to fix this issue, and they do this by predicting what information is important to remember, and what can be ignored. LSTMs were originally introduced in 1997 by Sepp Hochreiter and Jürgen Schmidhuber's paper *Long Short-Term Memory*. The structure of LSTMs since then has had some variations made to improve the general structure for specific tasks. In 2015 a group of researchers at Google tried to improve the generally excepted structure even further and set out to discover any possible improvements that could be made. Ultimately, in their paper *An Empirical Exploration of Recurrent Network Architectures* they found through a straightforward architecture search that the existing structure of LSTMs can't be beat for most applications.

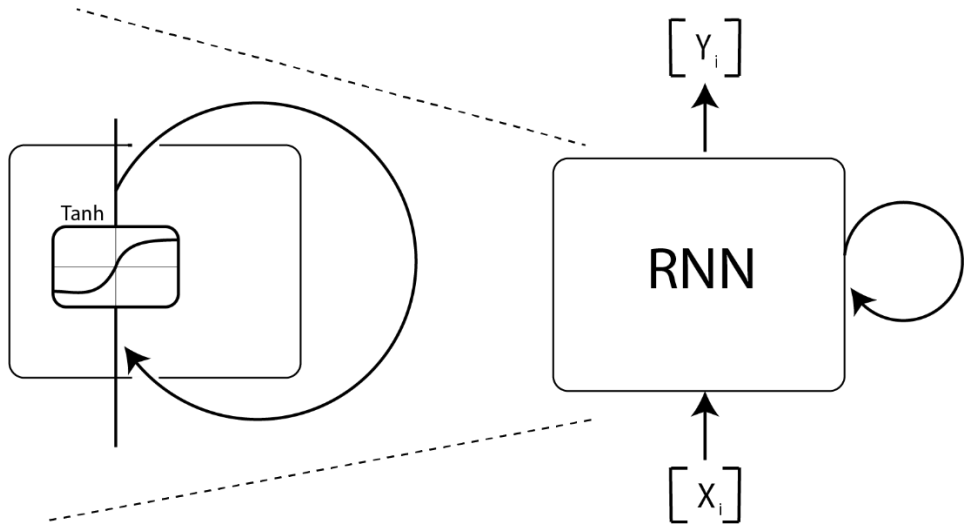

*The inner workings of an RNN cell*

A simple RNN has a single layer (a tanh layer in this diagram) that calculates the output using both the state of previous inputs and the current input.

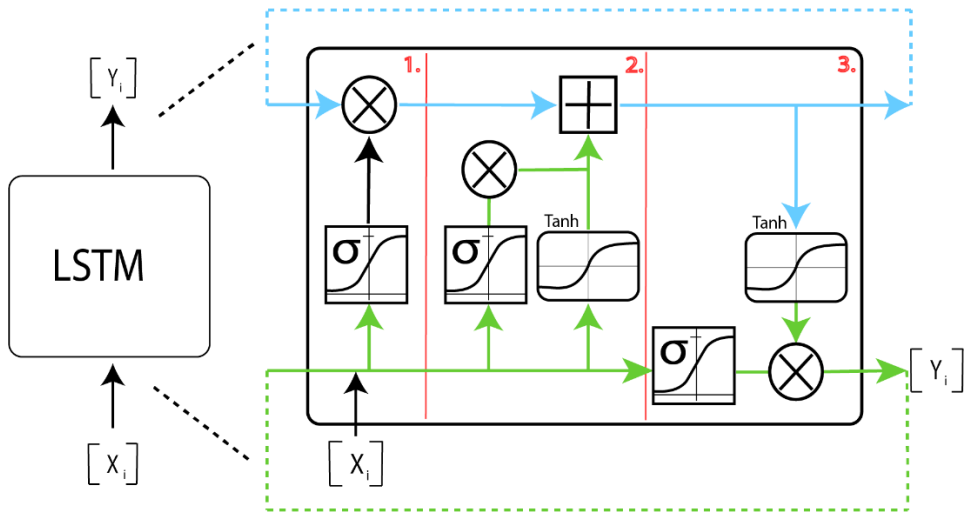

*High level overview of an LSTM cell*

An LSTM cell consists of two channels running from left to right: a top channel where information about state flows from one iteration to the next, and a bottom channel that uses the previous output to determine how to modify the state of the top channel. [4] The bottom channel changes the state of the top channel using so-called 'gates' to regulate the flow of information. The first gate (labeled as 1) is called the 'forget' gate, and it pointwise multiplies the cell state channel with a sigmoid activation layer to determine which information should be forgotten (multiplication by 1 means completely remember, and by 0 means completely forget). The second gate (labeled as 2) is called the 'remember' gate, and it uses two layers to decide what information in the cell state channel should be reinforced. The final part of an LSTM (labeled as 3) uses both the top and bottom channels to decide what the network should output. The output of both channels is fed back into the network. [3]

**2.3 I/O**

The songs in our data are a mix of classical music and video game soundtracks consisting of single-track piano music. These songs have been converted to a Musical Instrument Digital Interface (MIDI) file format, which is a standard for specifying musical instructions.

To convert the MIDI files in our dataset into an input format that our model can train on, we first utilized the Music21 library's converter to parse MIDI data into a numerical format. Then we created a sequential list of Music21 note objects. Music21 extracts a comprehensive amount of information, which could be used to further improve input for training; however, for our purposes, we use the note's pitch(s) (plurality for chords), whether the note's a rest, and the note's duration.

Once we have parsed information from the song, we generate two lists for each attribute of the notes we are using. The first lists are sequences the correspond to sequential notes in a song, and the second lists correspond to the note the comes immediately after the sequence in the first list. Finally, we one hot encoded all the lists to remove bias that would occur from assigning numeric values.

Since our network is a supervised model, we must provide a "correct" note/category that the model will compare its prediction with. In this case, this is the note immediately after each chunk which resides in the previously mentioned second list.

For our model to generate a prediction, we first take the pre-parsed notes data from our training data and take a random fifty note chunk and feed it to the model to predict the next five-hundred notes and durations. This output is then converted back into a sequence of Music21 note objects. Finally, the sequence of five-hundred generated notes is used to generate a MIDI file containing the model's predictions.

## 2.4 Outputs

The following score is an example of one of the outputs generated by the network. As can be seen, the network is able to utilize a large variety of notes, steps, and techniques throughout this portion of the score.

To generate the outputs after the network has created a sequence of notes, we used the same Music21 library as before but rather than extracting note data from midi files we created midi files from note data.

# 3 Experiments

To find which network structures and hyperparameters would yield the best results, we created a multitude of models with variations in hyperparameters such as batch size and learning rate, as well as the optimizations and activation functions, and the structure and size of our layers, then for each of those variations we generated a set of five songs which start predictions from the same input. Once this output is achieved, we evaluated each model on how we liked the songs. We then repeated the architecture search process on the best performing models. We take inspiration from evolutionary architecture and hyperparameter search in this process, however as music quality is a subjective metric without a trivial automatic validation metric there is no way that evolution can be done without human intervention.

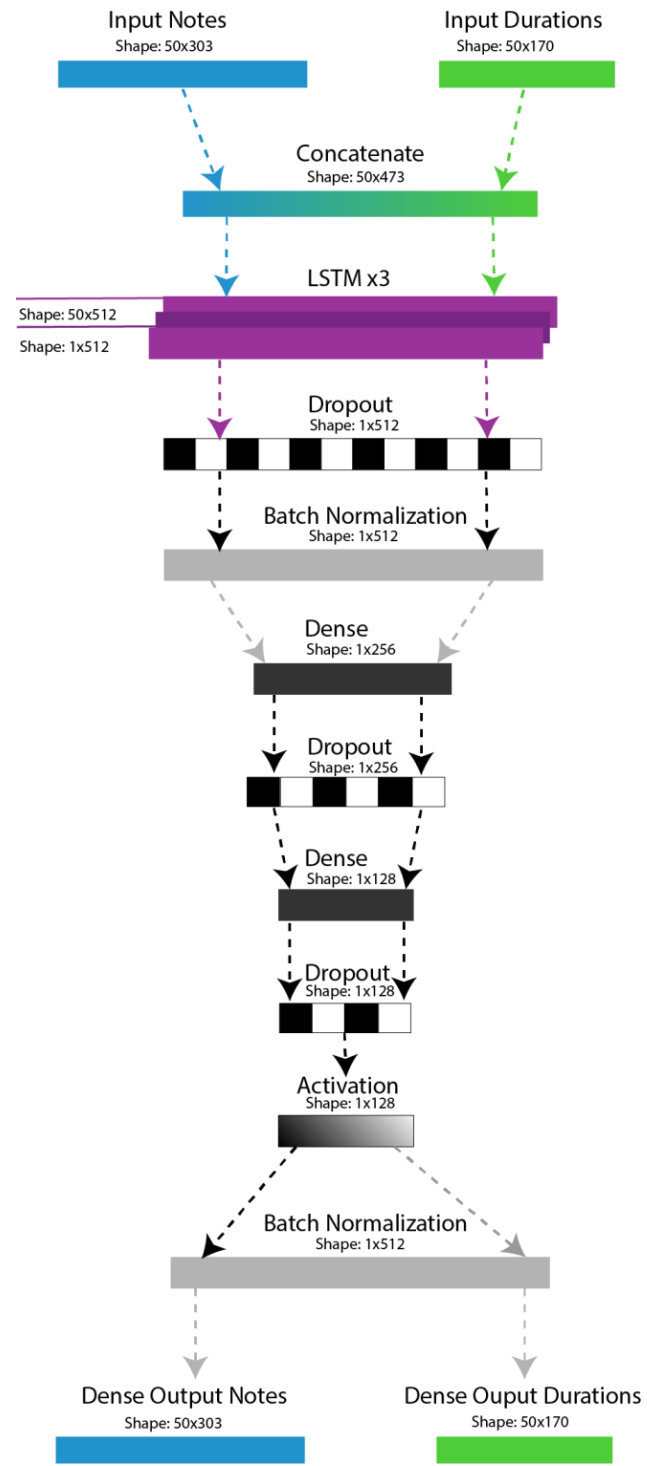

## 3.1 Input Structure

When creating the network, we experimented with different ways to represent our data, including representing notes with all their attributes as single integer inputs, transferring integer representations into floats between zero and one, as well as one hot encoding. In the end, we found the best results from our input was to separate each of the different attributes of each note into their own input layers and represent that data using one hot encoding and then later concatenate them together to be fed into the network.

We have experimented with training the inputs separately before concatenation, however this did not yield good results in any of our experiments, therefore we ultimately decided to drop this idea.

## 3.2 Network Structure

In experimenting with different types of network structures, we found that for our current input format, the following structure works the best. This structure built upon one used in an article for music generation similar to what we have done here (6). The major difference we found between our structure and others, was that with the large amount of processing power available to us via MSOE's high performance computing cluster named Rosie, we were able to make our network larger, and deeper with more stacked LSTMs at the front of the network. This allows the network to notice more patterns and characteristics that a shallower network would struggle to recognize. To better characterize the difference this deeper architecture gives, consider a network trained to classify photographs with three hidden layers. The first layer of the network may identify edges and contrast, the second layer may find where certain shapes are, and the third layer may decide on what real-world object those shapes represent. A shallower network would have to rely on less powerful abstractions to classify the image. In the same sense, our network's tiered LSTM architecture may allow it to pick apart the various aspects of music such as if it is building tension, speeding up, or looking at any other type of pattern that may be found in music.

## 4 Challenges

In the process of building and experimenting with this project, we ran into many challenges, many of which we have overcome, and others which we plan to address in future work. Among these includes how to evaluate non-deterministic outputs, how the time between notes should be represented, how the data as a whole should be represented, and how to have multiple tracks of music generated.

### 4.1 Past Challenges

As has been previously mentioned, music evaluation is subjective and cannot trivially be assessed without human interaction. While most people can listen to a melody and notice if there is a note suddenly played out of key, and it would be very reasonable for a computer to determine this as well, we all still have our own preferences on what sounds 'good'. When training a model to create music there is no definite way for it to really judge itself on if the music is good or not, this can make comparing models quite difficult. On top of this, depending on what sequence we give the models to complete a single model may produce amazing, or terrible outputs. In order to help mitigate this issue, we have each model generate a list of outputs to be judged by human evaluators.

In hand-written music, there are plenty of cases where the music may make sudden changes or jumps with reasoning known only to the songwriter. For instance, to build the tension a song may repeat the same note many times. These cases would confuse many of the models we made, and when generating their own music, models would tend to repeat the same notes forever if they repeated more than a few times.

### 4.2 Future Plans

In its present state, our models do not know how to create beginnings and endings to pieces, rather, it just starts playing, and ends abruptly at some designated number of notes. Creating a way for the model to dynamically decide when to end a piece would allow for non-abrupt endings.

Currently our models use and create single-track songs only. In the future, we will be allowing our models to create multiple tracks that work together, for instance, it may produce a melody with a bass to accompany it.

The current input of our model consists of notes and durations for notes. Any chords in the data set are considered unique notes for our model. This approach is quite limiting since it

1) Does not differentiate between notes and chords. To give an example, a C major triad is not seen as a composition of notes, but rather as its own unique note altogether. Allowing the network to learn how notes can form chords would allow it to learn a deeper understanding of music.
2) Does not allow some complex compositional patterns. Some examples are arpeggios, melodies over chords, and meaningful harmony.

In the future, we plan to use the time difference between note on and note off events, very akin to how MIDI is represented in .mid files. This would allow the model to compose music more like a human who thinks of chords as the composition of notes.

### 5 Conclusion

In this paper, we gave an overview of the evolution of LSTMs and how we applied them to music generation. Extending a previous work, we had the data formatted to use one hot encoding, gave an abundance of different types and genres of music for our network to learn on, and, rather than only predicting the notes of a song, we expanded our network to also predict the durations of notes as well. In the future, we will be expanding our network to learn and use even more attributes of a song, understand beginning and endings to songs, and allow for multi-track inputs and outputs.